\documentstyle[twoside]{article}

\catcode`\@=11
\long\def\@makefntext#1{
\protect\noindent \hbox to 3.2pt {\hskip-.9pt  
$^{{\eightrm\@thefnmark}}$\hfil}#1\hfill}		

\def\thefootnote{\fnsymbol{footnote}}
\def\@makefnmark{\hbox to 0pt{$^{\@thefnmark}$\hss}}	
	
\def\ps@myheadings{\let\@mkboth\@gobbletwo
\def\@oddhead{\hbox{}
\rightmark\hfil\eightrm\thepage}   
\def\@oddfoot{}\def\@evenhead{\eightrm\thepage\hfil
\leftmark\hbox{}}\def\@evenfoot{}
\def\sectionmark##1{}\def\subsectionmark##1{}}

\oddsidemargin=\evensidemargin
\renewcommand{\thefootnote}{\fnsymbol{footnote}}

\newcounter{sectionc}\newcounter{subsectionc}\newcounter{subsubsectionc}
\renewcommand{\section}[1] {\vspace{12pt}\addtocounter{sectionc}{1} 
\setcounter{subsectionc}{0}\setcounter{subsubsectionc}{0}\noindent 
	{\tenbf\thesectionc. #1}\par\vspace{5pt}}
\renewcommand{\subsection}[1] {\vspace{12pt}\addtocounter{subsectionc}{1} 
	\setcounter{subsubsectionc}{0}\noindent 
	{\bf\thesectionc.\thesubsectionc. {\kern1pt \bfit #1}}\par\vspace{5pt}}
\renewcommand{\subsubsection}[1] {\vspace{12pt}\addtocounter{subsubsectionc}{1}
	\noindent{\tenrm\thesectionc.\thesubsectionc.\thesubsubsectionc.
	{\kern1pt \tenit #1}}\par\vspace{5pt}}
\newcommand{\nonumsection}[1] {\vspace{12pt}\noindent{\tenbf #1}
	\par\vspace{5pt}}

\newcounter{appendixc}
\newcounter{subappendixc}[appendixc]
\newcounter{subsubappendixc}[subappendixc]
\renewcommand{\thesubappendixc}{\Alph{appendixc}.\arabic{subappendixc}}
\renewcommand{\thesubsubappendixc}
	{\Alph{appendixc}.\arabic{subappendixc}.\arabic{subsubappendixc}}

\renewcommand{\appendix}[1] {\vspace{12pt}
        \refstepcounter{appendixc}
        \setcounter{figure}{0}
        \setcounter{table}{0}
        \setcounter{lemma}{0}
        \setcounter{theorem}{0}
        \setcounter{corollary}{0}
        \setcounter{definition}{0}
        \setcounter{equation}{0}
        \renewcommand{\thefigure}{\Alph{appendixc}.\arabic{figure}}
        \renewcommand{\thetable}{\Alph{appendixc}.\arabic{table}}
        \renewcommand{\theappendixc}{\Alph{appendixc}}
        \renewcommand{\thelemma}{\Alph{appendixc}.\arabic{lemma}}
        \renewcommand{\thetheorem}{\Alph{appendixc}.\arabic{theorem}}
        \renewcommand{\thedefinition}{\Alph{appendixc}.\arabic{definition}}
        \renewcommand{\thecorollary}{\Alph{appendixc}.\arabic{corollary}}
        \renewcommand{\theequation}{\Alph{appendixc}.\arabic{equation}}
        \noindent{\tenbf Appendix \theappendixc #1}\par\vspace{5pt}}
\newcommand{\subappendix}[1] {\vspace{12pt}
        \refstepcounter{subappendixc}
        \noindent{\bf Appendix \thesubappendixc. {\kern1pt \bfit #1}}
	\par\vspace{5pt}}
\newcommand{\subsubappendix}[1] {\vspace{12pt}
        \refstepcounter{subsubappendixc}
        \noindent{\rm Appendix \thesubsubappendixc. {\kern1pt \tenit #1}}
	\par\vspace{5pt}}

\newcommand{\textlineskip}{\baselineskip=13pt}
\newcommand{\smalllineskip}{\baselineskip=10pt}

\def\eightcirc{
\begin{picture}(0,0)
\put(4.4,1.8){\circle{6.5}}
\end{picture}}
\def\eightcopyright{\eightcirc\kern2.7pt\hbox{\eightrm c}} 

\newcommand{\copyrightheading}[1]
	{\vspace*{-2.5cm}\smalllineskip{\flushleft
	{\footnotesize International Journal of Modern Physics A, #1}\\
	{\footnotesize $\eightcopyright$\, World Scientific Publishing
	 Company}\\
	 }}

\def\abstracts#1#2#3{{
	\centering{\begin{minipage}{4.5in}\baselineskip=10pt\footnotesize
	\parindent=0pt #1\par 
	\parindent=15pt #2\par
	\parindent=15pt #3
	\end{minipage}}\par}} 

\newcommand{\bibit}{\nineit}

\renewenvironment{thebibliography}[1]
	{\frenchspacing
	 \ninerm\baselineskip=11pt
	 \begin{list}{\arabic{enumi}.}
	{\usecounter{enumi}\setlength{\parsep}{0pt}
	 \setlength{\leftmargin 12.7pt}{\rightmargin 0pt} 
	 \setlength{\itemsep}{0pt} \settowidth
	{\labelwidth}{#1.}\sloppy}}{\end{list}}

\newcounter{itemlistc}
\newcounter{romanlistc}
\newcounter{alphlistc}
\newcounter{arabiclistc}

\newenvironment{romanlist}
	{\setcounter{romanlistc}{0}
	 \begin{list}{$($\roman{romanlistc}$)$}
	{\usecounter{romanlistc}
	 \setlength{\parsep}{0pt}
	 \setlength{\itemsep}{0pt}}}{\end{list}}

\newcommand{\fcaption}[1]{
        \refstepcounter{figure}
        \setbox\@tempboxa = \hbox{\footnotesize Fig.~\thefigure. #1}
        \ifdim \wd\@tempboxa > 5in
           {\begin{center}
        \parbox{5in}{\footnotesize\smalllineskip Fig.~\thefigure. #1}
            \end{center}}
        \else
             {\begin{center}
             {\footnotesize Fig.~\thefigure. #1}
              \end{center}}
        \fi}

\newcommand{\tcaption}[1]{
        \refstepcounter{table}
        \setbox\@tempboxa = \hbox{\footnotesize Table~\thetable. #1}
        \ifdim \wd\@tempboxa > 5in
           {\begin{center}
        \parbox{5in}{\footnotesize\smalllineskip Table~\thetable. #1}
            \end{center}}
        \else
             {\begin{center}
             {\footnotesize Table~\thetable. #1}
              \end{center}}
        \fi}

\def\@citex[#1]#2{\if@filesw\immediate\write\@auxout
	{\string\citation{#2}}\fi
\def\@citea{}\@cite{\@for\@citeb:=#2\do
	{\@citea\def\@citea{,}\@ifundefined
	{b@\@citeb}{{\bf ?}\@warning
	{Citation `\@citeb' on page \thepage \space undefined}}
	{\csname b@\@citeb\endcsname}}}{#1}}

\newif\if@cghi
\def\cite{\@cghitrue\@ifnextchar [{\@tempswatrue
	\@citex}{\@tempswafalse\@citex[]}}
\def\citelow{\@cghifalse\@ifnextchar [{\@tempswatrue
	\@citex}{\@tempswafalse\@citex[]}}
\def\@cite#1#2{{$\null^{#1}$\if@tempswa\typeout
	{IJCGA warning: optional citation argument 
	ignored: `#2'} \fi}}

\def\pmb#1{\setbox0=\hbox{#1}
	\kern-.025em\copy0\kern-\wd0
	\kern.05em\copy0\kern-\wd0
	\kern-.025em\raise.0433em\box0}

\def\fnt#1#2{\footnotetext{\kern-.3em
	{$^{\mbox{\scriptsize #1}}$}{#2}}}

\def\fpage#1{\begingroup
\voffset=.3in
\thispagestyle{empty}\begin{table}[b]\centerline{\footnotesize #1}
	\end{table}\endgroup}

\def\runninghead#1#2{\pagestyle{myheadings}
\markboth{{\protect\footnotesize\it{\quad #1}}\hfill}
{\hfill{\protect\footnotesize\it{#2\quad}}}}
\font\tenrm=cmr10
\font\tenit=cmti10 
\font\tenbf=cmbx10
\font\bfit=cmbxti10 at 10pt
\font\ninerm=cmr9
\font\nineit=cmti9

\font\eightrm=cmr8

\def\qed{\hbox{${\vcenter{\vbox{			
   \hrule height 0.4pt\hbox{\vrule width 0.4pt height 6pt
   \kern5pt\vrule width 0.4pt}\hrule height 0.4pt}}}$}}

\renewcommand{\thefootnote}{\fnsymbol{footnote}}	

\def\be{\begin{equation}}       \def\eq{\begin{equation}}
\def\ee{\end{equation}}         \def\eqe{\end{equation}}

\def\bea{\begin{eqnarray}}      \def\eqa{\begin{eqnarray}}
\def\ena{\end{eqnarray}}        \def\eea{\end{eqnarray}}
                                \def\eqae{\end{eqnarray}}

\def\ba{\begin{array}}
\def\ea{\end{array}}
\def\unit{1 \hskip-.3em \raise2pt\hbox{$ \scriptstyle |$ } }

\def\c{\gamma} 
\def\d{\delta}
\def\e{\epsilon}           
\def\f{\phi}               
\def\g{\gamma}

\def\k{\kappa}                    
\def\l{\lambda}
\def\m{\mu}
\def\n{\nu}
  
\def\p{\pi}                
  \def\th{\theta}                  
\def\r{\rho}                                     
\def\s{\sigma}                                   
\def\t{\tau}

\def\G{\Gamma}

\def\L{\Lambda}
\def\O{\Omega}

\def\ca{{\cal A}}

\def\co{{\cal O}}

\def\>{\rangle} 

\def\<{\langle} 

\def\bop#1{\setbox0=\hbox{$#1M$}\mkern1.5mu
        \vbox{\hrule height0pt depth.04\ht0
        \hbox{\vrule width.04\ht0 height.9\ht0 \kern.9\ht0
        \vrule width.04\ht0}\hrule height.04\ht0}\mkern1.5mu}

\def\g0{g_{(0)}}

\newcommand{\sm}[1]{\mbox{\scriptsize #1}} 
\newcommand{\tnnn}[1]{\mbox{\tiny #1}} 
\def\GN{G_{\mbox{\tnnn N}}}
\def\Tr{{\rm Tr}\, }

\def\half{{1 \over 2}}

\def\pa{\partial}

\def\nonu{\nonumber \\{}}
\def\half{{1 \over 2}}

\def\bfr{\begin{flushright}}
\def\efr{\end{flushright}}

\begin{document}

\runninghead{Asymptotically Anti-de Sitter Spacetimes
and ...} {Asymptotically Anti-de Sitter Spacetimes
and ...}

\normalsize\textlineskip
\thispagestyle{empty}
\setcounter{page}{1}

\copyrightheading{}			

\vspace*{0.88truein}

\fpage{1}
\centerline{\bf ASYMPTOTICALLY ANTI-DE SITTER SPACETIMES}
\vspace*{0.035truein}
\centerline{\bf AND THEIR STRESS ENERGY TENSOR}
\vspace*{0.37truein}
\centerline{\footnotesize KOSTAS SKENDERIS\footnote{
kostas@feynman.princeton.edu}}
\vspace*{0.015truein}
\centerline{\footnotesize\it Physics Department, 
Princeton University}
\baselineskip=10pt
\centerline{\footnotesize\it Princeton, NJ 08544, USA}

\vspace*{0.21truein}
\abstracts{We consider asymtotically anti-de Sitter spacetimes
in general dimensions. We review the origin of infrared
divergences in the on-shell gravitational action, and the 
construction of the renormalized on-shell action by the addition 
of boundary counterterms. In odd dimensions, the renormalized on-shell action 
is not invariant under bulk diffeomorphisms that yield conformal 
transformations in the boundary (holographic Weyl anomaly). 
We obtain formulae for the gravitational stress energy tensor, 
defined as the metric variation of the renormalized on-shell action, 
in terms of coefficients in the asymptotic expansion of the metric 
near infinity. The stress energy tensor transforms anomalously
under bulk diffeomorphisms broken by infrared divergences.
}{}{}


\vspace*{1pt}\textlineskip	
\section{Introduction} 
\vspace*{-0.5pt}
\noindent

\textheight=7.8truein
\setcounter{footnote}{0}
\renewcommand{\thefootnote}{\alph{footnote}}

Anti-de Sitter (AdS) spacetimes were studied in the eighties 
because they appear as ground states of many gauged supergravities.
In recent times they have attracted a lot of attention due to the 
AdS/CFT correspondence. The construction of conserved charges
for asymptotically AdS spaces was addressed in the eighties, 
see for instance Refs.\cite{AbDe}. Recent work appeared in
Ref.\cite{AshDas}. The AdS/CFT duality
has provided us with new insights and results about 
AdS gravity. It is the purpose of this contribution to extract these 
results and put them in a purely gravitational context. 

Our considerations will be at the classical level. We would however like  
to view our results as the lowest order approximation to the quantum theory.
The (logarithm of the) partition function of gravity is given 
to lowest order by the on-shell value of the gravitational action.
It is thus essential that the on-shell action is finite.
For asymptotically AdS solutions the on-shell action is proportional 
to the volume of spacetime and thus diverges since the spacetime 
is non-compact. 
One may deal with this divergence by regulating the on-shell action, 
computing all infinities, adding counterterms to cancel them and then 
removing the regulator. The resulting renormalized on-shell action
is manifestly finite. This procedure was carried out 
in Ref. \cite{HS}. 
  
The renormalized on-shell action does not preserve all symmetries
that the un-renormalized (and thus infinite) on-shell action formally
does. This is so because
some of the infrared divergences do not preserve all symmetries.
Adding counterterms to the action to cancel the infinities 
results in a renormalized on-shell action that transforms
anomalously under some of the original symmetries.
In particular, it was found in Ref.\cite{HS} that in odd dimensions 
the counterterms break the bulk diffeomorphisms that generate Weyl 
transformations on the boundary. This means that the renormalized 
on-shell action depends on the chosen metric on the boundary, 
not just on its conformal class. This anomaly corresponds, via the AdS/CFT 
correspondence, to the conformal anomaly of the dual CFT. 
We would like to emphasize that the anomaly does not 
depend on the specific way one chooses to do the computation. 
Any scheme that results in a finite  on-shell action for arbitrary 
solutions of Einstein's equations with negative cosmological constant
will have this property. 

Motivated by the AdS/CFT correspondence, it was proposed 
in Ref.\cite{BK} to define the stress energy tensor associated 
with asymptotically AdS spacetimes as the metric variation of the renormalized 
on-shell action. 
Explicit formulae for the stress energy tensor in terms of 
coefficients in the asymptotic expansion of the metric near 
infinity was obtained in Ref.\cite{KSS}
The stress energy tensor so defined does not transform covariantly under
all bulk diffeomorphisms because the on-shell renormalized action 
transforms anomalously. Notice that the energy momentum tensor 
may transform anomalously even if the anomaly in the action vanishes
for a given background. In other words, the value of the anomaly 
may vanish, but not the value of its metric variation. This case is 
encountered for global AdS. Only if the anomaly vanishes identically 
will the stress energy tensor transform covariantly. This 
is true for even dimensional asymptotically AdS spaces.
In Ref.\cite{AshDas} it was argued that the counterterm subtraction method 
is not covariant (in odd dimensions). Indeed, this is the case. 
The non-covariance, however, is forced upon us by infrared divergences 
in the on-shell action, and these divergences cannot be ignored.

This paper is organized as follows. In the next section we recall 
the definition of asymptotically AdS spacetimes. 
In section 3, we review the computation
of the infrared divergences and the construction of the on-shell
renormalized action. Section 4 contains the computation
of the stress energy tensor.

\section{Asymptotically AdS Spacetimes}
\noindent

Anti-de Sitter (AdS) spacetime is a maximally symmetric solution of 
Einstein's equations with negative cosmological constant,
\be \label{feq1}
R_{\m \n} - \half R G_{\m \n} = \L G_{\m \n},
\ee
where $\L$ is the negative cosmological constant. 
AdS spacetime is homogeneous and isotropic and
its isometry group is $SO(2,d)$. These conditions imply that 
the $AdS$ solution is conformally flat and satisfy the stronger equation,
\be \label{curv}
R_{\k \l \m \n} = l^2 (G_{\k \m} G_{\l \n} - G_{\k \n} G_{\l \m}),
\ee 
where $l$ is related to the cosmological constant as $\L=-d(d-1)/2l^2$.

The metric for AdS${}_{d+1}$ is given by
\be \label{met1}
ds^2 = l^2\left[-(1+r^2) dt^2 + {dr^2 \over (1+r^2)} + r^2 d\O_{d-1}\right].
\ee
At infinity the space
has a boundary with topology $R \times S^{d-1}$ (we actually consider the
covering space of AdS). 
Let us introduce a new coordinate $\tan \th=r$. The metric becomes
\be \label{met2}
ds^2 = {l^2 \over \cos^2 \th}\left[-dt^2 + d\th^2 + \sin^2 \th d\O_{p-1}
\right].
\ee
The boundary is now located at $\th=\pi/2$. The metric
has a second order pole at the boundary, so it does not induce a 
metric there. To obtain a metric one
picks a positive function $r$ with a single zero at the boundary,
and then evaluates $r^2 G$ at the boundary,
\be
g_{(0)} = r^2 G|_{R \times S^{d-1}}.
\ee
In our case, one can pick $r=\cos \th$. Then $g_{(0)}$ is the standard metric 
on a Lorentzian cylinder. The metric $g_{(0)}$, however, depends on the 
choice of defining function. Different defining functions
are related by $r'=r e^w$. It follows that $g_{(0)}$ is well-defined up to 
conformal transformation. Thus the AdS metric induces a conformal 
structure (i.e. a metric up to conformal transformations) at the boundary. 

To define asymptotically AdS spacetimes it is useful to recall
the notion of conformal infinity introduced by Penrose. 
Let $X$ be the interior of a manifold-with-boundary $\bar{X}$,
$M$ the (compact) boundary of $\bar{X}$, and $G$ a  
metric on $X$. The metric $G$ is said to be conformally compact 
if there exists a defining function $r$ (i.e. $r>0$ in $X$
and $r=0, dr \neq 0$ on $M$) such that 
$r^2 G$ can be smoothly extended as a metric on $\bar{X}$.
This procedure defines a conformal structure at the boundary 
in the way described in the previous paragraph. 
If in addition $G$ satisfies (\ref{feq1})
we will say that $X$ is an ``asymptotically $AdS$ spacetime''.
Notice that that this definition is more general than the one  
employed in Ref.\cite{AshDas} since we  
do not impose any restriction on the topology of the boundary. 

Given a conformal structure, can one obtain an
asymptotically AdS spacetime with this conformal structure 
at infinity? One may view this question as a Dirichlet 
problem for AdS gravity. This question has been investigated in the mathematics
literature. It was shown by Fefferman and Graham in 
Ref.\cite{FeffermanGraham} that one can obtain an asymptotic solution 
of Einstein's equations given a representative
of the conformal structure. The solution is of the form
\bea \label{coord}
&&ds^2=G_{\m \n} dx^\m dx^\n = l^2 \left({d\r^2 \over 4 \r^2} + 
{1 \over \r} g_{ij}(x,\r) dx^i dx^j \right), \nonu
&&g(x,\r)=g_{(0)} + \cdots + \r^{d/2} g_{(d)} + h_{(d)} \r^{d/2} \log \r + ... 
\eea
Several comments are in order here:
\begin{romanlist}
\item Any asymptotically AdS metric can be brought into this form
near the boundary. The boundary is located at $\r=0$. 
\item Einstein's equations can be solved order by order in the 
$r$ variable. Given a metric $g_{(0)}$ one can uniquely determine
the coefficients $g_{(2)},...,g_{(d-2)}$ and $h_{(d)}$ in terms 
of $g_{(0)}$.
\item The coefficient $h_{(d)}$ is present only when $d$ is even.
It is equal to the metric variation of the coefficient 
of the infrared logarithmic divergence in 
the on-shell gravitational action (discussed below). 
It is traceless and covariantly conserved. 
\item The asymptotic analysis determines only the trace 
and covariant divergence of $g_{(d)}$.
\end{romanlist}
The explicit expression for $g_{(2)},...,g_{(d-2)},h_{(d)}$ 
and the equations for the trace and covariant divergence of $g_{(d)}$ 
can be found in appendix A of Ref.\cite{KSS}
 
A case where the Dirichlet problem can be solved exactly
is when the bulk metric is conformally flat.\cite{SS}
In this case one obtains an exact solution given by the metric (\ref{coord}) 
with 
\bea\label{KS}
&&g(x,\rho )=g_{(0)}(x)+g_{(2)}(x)\rho+g_{(4)}(x)\rho^2~~, \nonu
&&g_{(2)ij}={1 \over d-2}(R_{ij} - {1 \over 2(d-1)} R g_{(0)ij})~~,~~g_{(4)}
={1\over 4}(g_{(2)})^2~~,
\eea
where the curvatures are of the metric $g_{(0)}$.\footnote{
The formula for $g_{(2)}$ 
is valid for $d \neq 2$. The $d=2$ case is also covered in Ref.\cite{SS}} 

So far we have discussed the construction of a bulk metric given 
a representative of a conformal structure. One may ask how these
results change if one picks a different representative 
of the given  conformal class, i.e. how the coefficients 
$g_{(i)}$ transform if we let $g_{(0)} \to e^{2 \s(x)} g_{(0)}$. 
One may use the explicit formulae of $g_{(i)}$
in terms of $g_{(0)}$ in order to obtain these transformation rules.
An alternative way to determine them is to note that there are 
bulk diffeomorphism that preserve the form of the metric (\ref{coord})
and induce the transformation 
$g_{(0)} \to e^{2 \s(x)} g_{(0)}$. The infinitesimal
form of these bulk diffeomorphisms has been worked out in Ref.\cite{ISTY}
Here we discuss the corresponding finite transformations.
Consider the coordinate transformation
\be \label{transform}
\rho=\rho' e^{-2 \s(x')} + \sum_{k=2} a_{(k)}(x') \r'^k, \qquad
x^i= x'^i + \sum_{k=1} a_{(k)}^i(x') \r'^k.
\ee 
The requirement that the transformation leaves the form 
of the metric invariant uniquely fixes the coefficients 
$a_{(i)}$ and $a^i_{(i)}$.
The first few are given by
\bea \label{arule}
&&a_{(2)}=-\half(\pa \s)^2 e^{-4 \s}, \qquad 
a_{(3)}={1 \over 4} e^{-6 \s} 
\left( {3 \over 4} (\pa \s)^2 + \pa^i \s \pa^j \s g_{(2)ij} \right), \\
&&a^i_{(1)}=\half \pa^i \s e^{-2 \s}, \qquad
a^i_{(2)}=-{1 \over 4} e^{-4 \s} \left(\pa_k \s g_{(2)}^{ik} 
+ \half \pa^i \s (\pa \s)^2 + \half \G_{kl}^i \pa^k \s \pa^l \s \right),
\nonumber
\eea
where indices are raised and lowered with $g_{(0)}$.
With these results one can calculate the transformation rules of
$g_{(i)}$:
\bea \label{rules}
g_{(0)ij}' &=& e^{2 \s} g_{(0)ij} \nonu
g_{(2)ij}' &=& g_{(2)ij} + \nabla_i \nabla_j \s 
- \nabla_i \s \nabla_j \s + \half (\nabla \s)^2 g_{(0)ij} \nonu
g_{(4)ij}' &=& e^{-2 \s} \left[g_{(4)ij} -2 \s h_{(4)ij} 
-{1 \over 4} \nabla^k \s (\nabla_i g_{(2)jk} +\nabla_j g_{(2)ik}
-2 \nabla_k g_{(2)ij})
\right. \nonu
&+& {1 \over 4} (\nabla_i \nabla^k \s g_{(2)kj} 
+ \nabla_j \nabla^k \s g_{(2)ki})
+{1 \over 4} R_{kilj} \nabla^k \s \nabla^l \s
+ {1 \over 4} \nabla_i \nabla_j \s (\nabla \s)^2 \nonu
&+&({1 \over 16} 
(\nabla \s)^4 - {1 \over 4}\nabla^k \s \nabla^l \s g_{(2)kl}) g_{(0) ij}
+{1 \over 4} \nabla_i \nabla_k \s \nabla_j \nabla^k \s \nonu
&-&\left. {1 \over 8} 
(\nabla_i (\nabla \s)^2 \nabla_j \s+\nabla_j (\nabla \s)^2 \nabla_i \s)
 \right]. \nonu
g_{(d)ij}'&=&e^{-(d-2)\s} g_{(d)ij}, \qquad d=2k+1.
\eea  
The infinitesimal version of (\ref{arule}) and (\ref{rules}) agree
with the infinitesimal ones derived in Ref.\cite{ISTY}. 

\section{Infrared Divergences}
\noindent

We have argued in the previous section that asymptotically AdS 
spacetimes come equipped with a bulk Einstein metric and a corresponding
boundary conformal structure. We will now show that  
infrared divergences in the on-shell value of the action 
force a dependence on the boundary metric, not just on its conformal
class.

The gravitational equations (\ref{feq1}) can be derived from the 
action
\be \label{action}
S[G]={1 \over 16 \p \GN}[\int_{X} d^{d+1}x\, 
\sqrt{G}\, (R[G] + 2 \L) 
- \int_{\pa X} d^d x\, \sqrt{\c}\, 2 K],
\ee
where $K$ is the trace of the second fundamental form and
$\c$ is the induced metric on the boundary. 

The on-shell gravitational action diverges because of the infinite 
bulk volume and the fact that the induced metric diverges at the boundary.
To regulate the theory we cutoff the radial integration, $\r \geq \e$,
and evaluate the boundary term at $\r=\e$, where $\e$ is the regulator.
This regularization procedure was proposed in Ref.\cite{wit} and implemented 
in Ref.\cite{HS}

Using the asymptotic solution discussed in the previous section
one can evaluate the on-shell action. The result is \cite{HS}
\bea \label{regaction}
S_{\sm{reg}}&=&{1 \over 16 \p \GN}\left[\int_{\r\geq\e} 
d^{d+1}x\, \sqrt{G} \,(R[G] + 2 \L) 
- \int_{\r=\e} d^d x \sqrt{\c}\, 2 K\right] \nonu
&=& {1 \over 16 \pi \GN} \int d^d x \sqrt{\det g_{(0)}} \left( 
\epsilon^{-d/2} a_{(0)} + \epsilon^{-d/2+1} a_{(2)} + \ldots 
+ \epsilon^{-1} a_{(d - 2)} \right. \nonu
&& \left. \hspace{4cm} - \log \epsilon\, a_{(d)} \right) + \co(\e^0),
\eea
where the coefficients $a_{(n)}$ are local covariant expressions
of the metric $g_{(0)}$ and its curvature tensor. 
The explicit expressions can be found in appendix B of Ref.\cite{KSS}
Here we only give the coefficients of the logarithmic
divergences in $d=2$ and $d=4$: $a_{(2)}=\half R$ and 
$a_{(4)}=-R^{ij} R_{ij}/8 + R^2/24$.

This computation is universal and applies to any asymptotically AdS spacetime.
It is simple to show that the infrared divergences depend only on
the coefficients $g_{(2)},...,g_{(d-2)}$ in the asymptotic expansion of 
the metric. As we have seen in the previous section, these
coefficients are universal in the sense that they uniquely determined 
in terms of $g_{(0)}$. The result in (\ref{regaction}) does depend 
on the chosen regularization.
The power law divergences are regularization dependent. For instance,
a manifestly supersymmetric regularization will yield $a_{(0)}=0$.
The logarithmic divergence, however, is regularization independent.

We now proceed to renormalize the on-shell gravitational action
by adding counterterms to cancel the infrared divergences.\cite{HS}
Using minimal subtraction, we obtain for the renormalized action 
\bea \label{renaction}
S_{\sm{ren}}[g_{(0)}]&=&\lim_{\e \to 0}
\left(S_{\sm{reg}} -{1 \over 16 \p \GN} \int_{\r=\e} 
\sqrt{\c}[2(1-d) + {1 \over d-2} R \right. \nonu
&-& \left.{1 \over (d-4) (d-2)^2}
(R_{ij} R^{ij} - {d \over 4 (d-1)} R^2) - \log \e\, a_{(d)}] \right),
\eea
where we have written the counterterms in terms of fields on the 
regulating hypersurface $\r = \e$, as in ref.\cite{BK}. 
This formula should be understood as containing only divergent 
counterterms in each dimension. This means that in even dimension
$d=2k$ one should include only the first $k$ counterterms 
and the logarithmic one. In odd $d=2k+1$, only the first 
$k+1$ counterterms should be included. The logarithmic counterterms
appear only for $d$ even. The renormalized action (\ref{renaction})
is finite up to $d=6$. It is straightforward
but tedious to compute the necessary counterterms for $d>6$.
As we have remarked above, only the logarithmic term has an 
invariant meaning. The other counterterms are regularization 
and scheme dependent. In particular one can add further boundary 
terms provided they do not diverge. We will make use of this 
freedom in the next section. 
The logarithmic term was incorrectly (and surprisingly) 
omitted in a large part of the literature on the subject. 
Because of this counterterm the on-shell action depends on the 
chosen representative of the boundary conformal structure,
i.e. 
\be \label{holano}
S_{\sm{ren}}[e^{2 \s} g_{(0)}] = S_{\sm{ren}}[g_{(0)}] + \ca[g_{(0)},\s].
\ee
The anomalous term has been computed for infinitesimal $\s$
in Ref.\cite{HS} It is proportional to $a_{(d)}$.
This  anomalous transformation has been called 
the holographic Weyl anomaly because it corresponds 
to the Weyl anomaly of the corresponding CFT in the 
AdS/CFT correspondence.
Notice that in even dimensions $a_{(2k+1)}$ vanishes identically, so in 
these cases the on-shell action depends only on the conformal
class of the boundary metric.
This is in agreement with the fact that conformal field theories
in odd dimensions do not have conformal anomalies.

As we have discussed in the previous section, 
boundary conformal transformations are induced by
a specific class of bulk diffeomorphisms. It follows that the 
finite on-shell action in odd dimensions is {\em not} invariant 
under these diffeomorphisms. In other words, infrared divergences
break part of bulk diffeomorphisms. Only the bulk diffeomorphisms
that do not yield a boundary Weyl transformation are true
symmetries. The anomaly itself is a conformal invariant,
so its vanishing depends only on the conformal class of the 
boundary metric. 
It may vanish in one background, but its metric variation may not. 
This means that in backgrounds where 
the anomaly vanishes non-trivially, so the on-shell action 
is invariant under all diffeomorphisms, the stress energy 
tensor may still have an anomalous variation. This is 
exactly what happens for (globally) AdS spacetimes.

\section{Stress Energy Tensor}
\noindent

In the previous section we have obtained the on-shell value of the 
gravitational action as a functional of the boundary metric.
In Ref.\cite{BrownYork}, Brown and York proposed defining the 
gravitational energy momentum tensor through a Hamilton-Jacobi
analysis as the functional derivative of the on-shell action 
with respect to the boundary metric. In Ref.\cite{BK} 
Balasubramanian and Kraus considered the Brown-York energy 
momentum tensor for asymptotically AdS spaces. 
Including the counterterms of Ref.\cite{HS} (except for the 
logarithmic one), they showed that the resulting energy 
momentum tensor was finite in the cases they considered.
Their analysis was extended in Ref.\cite{KSS}, where completely
explicit expressions for the energy momentum tensor 
in any even dimension and up seven dimensions were obtained,
as we now review.

The stress energy tensor is defined to be 
\be \label{tij1}
T_{ij} = {2 \over \sqrt{\det \g0}} 
{\d S_{\sm{ren}} \over \d g_{(0)}^{ij}}.
\ee
This can be evaluated by first computing the energy momentum 
tensor in the regulated theory 
and then removing the regulator, 
\be \label{tij2}
T_{ij}=\lim_{\e \to 0} 
{2 \over \sqrt{\det g(x, \e)}} {\d S_{\sm{ren}} \over \d g^{ij}(x,\e)}  
=\lim_{\e \to 0}\left( {1 \over \e^{d/2-1}}\, T_{ij}[\c]\right),
\ee
where $T_{ij}[\c]$ is the stress energy tensor in terms 
of the induced metric $\c$ of the hypersurface $\r=\e$.
It receives a contribution $T^{{\sm reg}}[\c]$ from the original (regulated) 
action (\ref{regaction}), and a contribution $T^{{\sm ct}}[\c]$
from the counterterms. The former is equal to 
$T^{{\sm reg}}[\c]=-{1 \over 8 \p \GN}(K_{ij} - K \c_{ij})$; the latter 
can be easily calculated from the explicit form of the counterterms.

Equipped with the explicit asymptotic solutions one can now 
evaluate (\ref{tij2}). This is a rather tedious 
exercise. The details can be found in Ref.\cite{KSS}.
Here we only give the final result,
\be \label{tfin}
T_{ij}={d l \over 16 \p \GN}\, (g_{(d)ij} + X_{ij}^{(d)}),
\ee
where $X_{ij}^{(d)}$ depends on the dimension. The result for 
all odd $d$, and even $d$ up to six are:
\bea
&&X^{(2k+1)}_{ij}=0, \qquad
X^{(2)}_{ij} =  - g_{(0)ij}\,\Tr\, g_{(2)} \nonu
&&X_{ij}^{(4)}=-{1 \over 8} g_{(0)ij} [(\Tr\, g_{(2)})^2-\Tr\, g_{(2)}^2] 
-\half (g_{(2)}^2)_{ij} + {1 \over 4} g_{(2)ij} \Tr\, g_{(2)} \nonu
&&X^{(6)}_{ij}=-A_{(6) ij} + S_{ij}
\eea
where 
\bea
\label{X}
&&\hspace{-0.6cm}A_{(6) ij}= {1 \over 3} \left(
2(g_{(2)} g_{(4)})_{ij}+(g_{(4)} g_{(2)})_{ij}-(g_{(2)}^3)_{ij} 
+ {1 \over 8}\,[\Tr\, g_{(2)}^2 - (\Tr\, g_{(2)})^2]\, g_{(2) ij} \right. \nonu
&&\hspace{0.6cm}-\Tr\, g_{(2)}\,[g_{(4)ij} - \half (g_{(2)}^2)_{ij}] 
-[{1 \over 8} \Tr\, g_{(2)}^2 \Tr\, g_{(2)} - {1 \over 24} (\Tr\, g_{(2)})^3
\nonu
&&\hspace{0.6cm}\left.-{1 \over 6} \Tr\, g_{(2)}^3
+{1\over 2} \Tr \, (g_{(2)}g_{(4)})]\,g_{(0) ij} \right)\, \\
&&\hspace{-0.6cm}
S_{ij}={1 \over 24} \left(\nabla^2C_{ij}-2R^{k \ l}_{\ i \ j} C_{kl}
+4(g_{(2)}g_{(4)}-g_{(4)}g_{(2)})_{ij}
+{1\over 10}(\nabla_i\nabla_jB
-g_{(0)ij}\nabla^2 B) \right.\nonumber \\
&&\hspace{1cm} \left.
+{2\over 5}g_{(2)ij}B+g_{(0)ij}(-{2\over 3}\Tr \, g_{(2)}^3
-{4\over 15}(\Tr \,g_{(2)})^3+
{3\over 5}\Tr \, g_{(2)}\Tr \, g^2_{(2)}) \right), \nonu
&&\hspace{-0.6cm}
C_{ij}=(g_{(4)}-{1\over 2}g^2_{(2)}+{1\over 4}g_{(2)}\Tr \,g_{(2)})_{ij}+
{1\over 8}g_{(0)ij}B~~,~~
B=\Tr \, g^2_2-(\Tr \, g_2)^2~~. \nonumber
\eea
The vanishing of $X^{(2k+1)}_{ij}$ reflects the fact that the holographic 
Weyl anomaly identically vanishes in even spacetimes. In the computation
we used the freedom to add finite counterterms in order to  
to remove terms proportional to $h_{(d)}$. We remind the reader 
that $h_{(d)}$ is proportional to the metric variation of $a_{(d)}$
so by adding a finite counterterm proportional to $a_{(d)}$ 
one can remove all $h_{(d)}$ dependence from the energy momentum tensor.
 
One can check by explicit computation that the stress energy tensors 
so obtained are covariantly conserved and their trace correctly reproduces 
the holographic Weyl anomaly.

The stress energy transforms anomalously under bulk diffeomorphisms that
induce Weyl transformations in the boundary. Indeed, using the 
definition (\ref{tij1}) and the transformation of the on-shell
action one gets,
\be
T_{ij}[e^{2 \s} g_{(0)}]= e^{-(d-2) \s} \left(T_{ij}[g_{(0)}] + 
{1 \over \sqrt{g_{(0)}}} {\d \ca[g_{(0)}] \over \d g_{(0)}^{ij}}\right).
\ee
From here it follows that the stress energy tensor may transform 
anomalously even if $\ca$ evaluated on a particular
background is equal to zero. An alternative way to 
compute the dependence of the stress energy tensor on a given 
representative of the boundary conformal structure is 
to use the transformation rules of 
$g_{(i)}$ in (\ref{rules}) and formula (\ref{tfin})-(\ref{X}).
The infinitesimal transformations can be found in Ref. \cite{KSS}.

To summarize: given a solution of Einstein's equation with negative 
cosmological constant one can obtain the corresponding energy 
momentum tensor by first reaching the coordinate system (\ref{coord})
near the boundary, and then plugging in the coefficients 
$g_{(i)}$ in (\ref{tfin}). 

We finish this section with a few comments on the addition of matter 
fields. As in the case of pure gravity one first needs to 
obtain asymptotic solutions of the coupled gravity-matter system.
It can happen that the leading behavior of the 
matter stress energy tensor is more singular than the leading 
behavior of the Einstein tensor. In this case the methods 
discussed here apply only if the fields parameterizing 
the matter boundary conditions are considered infinitesimal.
In the other cases one can proceed as in the case of pure gravity 
to regularize, renormalize the theory and obtain the stress
energy tensor as the metric variation of the on-shell action. 
A more detailed discussion can be found in Ref.\cite{KSS}

\subsection{Example: AdS$_5$}
\noindent
 
As an example we discuss in some detail the case of 
AdS${}_5$ spacetime. The coordinate transformation that brings
the metric (\ref{met1}) to the coordinate system in (\ref{coord}) is 
\be
r^2={1 \over \r} (1 - {\r \over 4})^2.
\ee
The metric coefficients are given by 
$g_{(0)} ={\rm diag}(-1,1,\sin^2 \th, \sin^2 \th \sin^2 \f)$,
$g_{(2)} =-\half {\rm diag}(1,1,\sin^2 \th, \sin^2 \th \sin^2 \f)$,
and $g_{(4)}={1 \over 16} g_{(0)}.$
The expansion terminates at $\r^2$ because AdS$_5$ 
is conformally flat.\cite{SS}
The expressions for $g_{(2)}$ and $g_{(4)}$ agree with formula (\ref{KS}).
Using (\ref{tfin}) we now get 
\be
T_{ij}= {l^3 \over 64 \p \GN}(4 \d_{i,0} \d_{j,0} + g_{(0)ij}) 
\ee
One can explicitly check that this stress energy tensor is conserved
and traceless. It is traceless because the conformal anomaly 
evaluated for global AdS vanishes. 

The boundary metric is conformally flat. The coordinate transformation 
\be \label{ctr}
\t \pm r = \tan \half (t \pm \th)
\ee
brings the metric to the form
\be
ds_{(0)}^2=4 \cos^2 \half (t+\th) \cos^2 \half (t-\th) 
\left( -d \t^2 + dr^2 + r^2 d\O_2 \right)
\ee
We can implement this Weyl transformation using the bulk 
diffeomorphism (\ref{transform}) with 
$e^{-2\s} = 4 \cos^2 \half (t+\th) \cos^2 \half (t-\th)$.
One can obtain the new metric by working out (\ref{rules}).
After some algebra one obtains $g_{(2)}'=g_{(4)}'=0$.
Alternatively, one can notice that the Dirichlet boundary problem
for conformally flat bulk metrics has a unique answer. In our case,
after the transformation $g_{(0)}'$ is flat, and from 
(\ref{KS}) we get $g_{(2)}'=g_{(4)}'=0$. 
It follows that $T_{ij}$ vanishes identically.

We thus see explicitly that the energy momentum tensor,
defined as the metric variation of the renormalized on-shell 
action, is {\em not} covariant with respect to diffeomorphisms 
that yield a Weyl transformation in the boundary. The anomalous 
transformation has its origin in the infrared divergences in the 
computation of the on-shell action. 

Notice that these results are in exact agreement with expectations 
from the AdS/CFT correspondence. The gravitational energy momentum
tensor is identified with the expectation value of the CFT 
stress energy tensor. This expectation value is zero for a CFT 
on flat space. When we consider the CFT on $R \times S^3$, however,
the expectation value is non-zero due to the Casimir energy.
Using the AdS/CFT dictionary for the gravity/gauge theory parameters
one finds exact agreement \cite{BK}. One could infer this 
agreement from the 
fact that the Casimir energy follows from the trace anomaly, and 
the latter was shown to be reproduced exactly by a gravity computation 
in Ref.\cite{HS}. These quantities had to agree, even though 
the gravity computation is at strong coupling and the CFT one in
weak coupling, because there is a non-renormalization theorem 
that protects them. 
 
To shed some more light on the results for the stress energy tensor,
one can work out the relation between the 
generator $H=\pa/\pa t$ of global time translations 
to generators of isometries of AdS with flat boundary.  
Using the coordinate transformation (\ref{ctr}), one obtains
\be
H=\half (P_\t + K_\t),
\ee
where $P_\t=\pa/\pa \t$ is the generator of $\t$-time translations and 
$K_i=x^2 \pa_i -2 x_i x^i \pa_j$ is the generator
of special conformal transformations. In a theory with conformal
anomalies, the dilatations and special conformal transformations
are broken. This allows for a non-zero eigenvalue of the generator
of global time translations acting on the ground state. 
Notice that in the case of Euclidean AdS global time
translations are mapped to dilatations, which are also broken.
 
\nonumsection{Acknowledgements}
\noindent
This research is supported in part by the NSF grant PHY-9802484.

\nonumsection{References}

\end{document}